\documentclass[aps,prb,twocolumn  ,superscriptaddress,nofootinbib,notitlepage]{revtex4-1}
\usepackage{graphicx}
\usepackage{amsmath}
\usepackage{amstext}
\usepackage{amssymb}
\usepackage{xfrac}
\usepackage[colorlinks,citecolor=blue]{hyperref}
\usepackage{graphicx}
\usepackage{amsmath}
\usepackage{amstext}
\usepackage{amssymb}
\usepackage{amsfonts}
\usepackage{longtable,booktabs}
\usepackage{hyperref}\usepackage{url}
\usepackage{subfigure}%
\usepackage{dsfont}

\usepackage{amsbsy}
\usepackage{dcolumn}
\usepackage{amsthm}
\usepackage{bm}
\usepackage{esint}
\usepackage{multirow}
\usepackage{hyperref}
\hypersetup{
    colorlinks=true,
    linkcolor=blue,
    filecolor=magenta,
    urlcolor=cyan,
}
\usepackage{cleveref}

\usepackage{mathrsfs}
\usepackage{amsfonts}
\usepackage{amsbsy}
\usepackage{dcolumn}
\usepackage{bm}
\usepackage{multirow}
\usepackage{color}

\newcommand{\comments}[1]{}

\def\Z{\mathbb{Z}}

  \usepackage{extarrows}
 
 \usepackage{datetime}

\begin{document}

  \title{\textbf{Fractional $S$-duality, Classification of Fractional Topological Insulators and Surface Topological Order}}

\author{Peng Ye}
  \affiliation{Department of Physics and Institute for Condensed Matter Theory, University of Illinois at Urbana-Champaign, IL 61801, USA}

\author{Meng Cheng}
  \affiliation{Department of Physics, Yale University, New Haven, CT 06520-8120, USA}

\author{Eduardo Fradkin}
  \affiliation{Department of Physics and Institute for Condensed Matter Theory, University of Illinois at Urbana-Champaign, IL 61801, USA}

    \begin{abstract}  
		In this paper,   we propose a generalization of the    $S$-duality  of four-dimensional quantum electrodynamics ($\text{QED}_4$) to $\text{QED}_4$ with fractionally charged excitations, the fractional $S$-duality. Such $\text{QED}_4$ can be obtained by gauging the $\text{U(1)}$ symmetry of a topologically ordered state with fractional charges. When time-reversal symmetry is imposed,  the axion angle ($\theta$) can take  a nontrivial but still time-reversal invariant value $\pi/t^2$ ($t\in\mathbb{Z}$). Here, $1/t$ specifies the minimal electric charge carried by bulk excitations. Such states  with time-reversal and $\text{U(1)}$ global symmetry (fermion number conservation) are fractional topological insulators (FTI). We propose a topological quantum field theory description, which microscopically justifies the fractional  $S$-duality.  Then, we consider stacking operations (i.e., a direct sum of Hamiltonians) among FTIs.  We find that there are two topologically   distinct classes of FTIs: type-I and type-II. Type-I ($t\in\mathbb{Z}_{\rm odd}$)   can be obtained by directly stacking a non-interacting topological insulator and a fractionalized gapped fermionic state with minimal charge $1/t$ and vanishing $\theta$. But type-II ($t\in\mathbb{Z}_{\rm even}$)   cannot be realized through any stacking.   Finally, we study the \emph{Surface Topological Order} of fractional topological insulators. 
   \end{abstract}
 \maketitle
\tableofcontents 

 \section{{Introduction}}
 
 Topological insulators (TI) \cite{TI4,TI6} in three dimensions (3D) are  {non-interacting} fermionic gapped states with   $\text{U(1)}$ symmetry and time-reversal   symmetry ($\mathsf{T}$).  The electromagnetic response of a TI \cite{Qi2008,lapa17,witten_RMP} is described by the effective action:
\begin{align}
S=i\frac{\theta}{8\pi^2}\int F \wedge  F+\frac{1}{2g^2}\int F\wedge \star F\,,\label{intro_action_main}
\end{align}
where the ``axion  angle'' $\theta=\pi\,\text{mod}\,2\pi$ in the TI bulk and $\theta=0\,\text{mod}\,2\pi$ in the vacuum or trivial insulators.    $F=dA$ is the 2-form field strength tensor of the external electromagnetic field $A_\mu$;    $g$ is the electromagnetic coupling constant.  
The presence of the $\theta$ term  leads to   the Witten effect \cite{Qi2008,witten1}: an externally inserted monopole  induces a polarization charge $\frac{\theta}{2\pi}$ localized on the monopole.  If $A$ is regarded as a dynamical gauge field, the action (\ref{intro_action_main}) describes a 4D quantum electrodynamics (QED$_4$) with an axion angle. It is known that this  theory has an ``$S$-duality'' \cite{Witten1995,VAFA19943,mo}. More precisely, the partition function  is invariant under the transformations $\mathcal{S}$ and $\mathcal{T}$:
 \begin{align}
	 \mathcal{S}:\,\, \tau\rightarrow -\frac{1}{\tau}\,\,;\,\,~~~~~~~~\mathcal{T}:\,\,\tau\rightarrow \tau+{1}\,,\label{eq_sec_intro_000}
 \end{align}
where the modular parameter $\tau $ is defined as $\tau=\frac{\theta}{2\pi}+i\frac{2\pi}{g^2}$.
$S$-duality has its origins in high energy physics and has been recently applied to new problems in condensed matter physics. Specifically, a new set of duality transformations have been proposed recently, that generalize the well-known particle-vortex duality \cite{son,metlitski_duality,senthil_duality,lesik_duality,shamit_duality,vortexduality1,vortexduality2}. One way to ``derive'' the duality is to consider the $S$-duality of QED$_4$ on an  open spacetime \cite{fradkin_kivelson_96,vortexduality1,bdr1,bdr2}.

In the presence of strong interactions, the concept of TI can be naturally generalized in two ways. One is the strongly-interacting topological insulator state with unfractionalized bulk as well as gapped, symmetric topologically-ordered surface states \cite{bti1,STO1,STO2,STO3,STO4,bti2}. The other direction is the fractional topological insulator (FTI) state whose bulk supports fractionalized excitations. In addition to bulk electron fractionalization, the most peculiar feature of FTIs is that $\theta/\pi$ is fractional and still consistent with $\mathsf{T}$.   As examples of 3D symmetry-enriched topological phases \cite{3dset_cheng,3dset_chen,3dset_plus,3dset_twisted,bti6,YW13a},   FTIs have been discussed  through parton constructions and solvable lattice models \cite{maciejkoFTI,maciejko_model,maciejko2015,swingle2011,ye16a,YW13a,2017arXiv170108828S,levin_lattice,swingle_fti_2012}.  However, using constructions alone it is not possible to  determine the values of all legitimate axion angles.
Especially, in parton constructions, there are infinitely many ways to  \emph{partition} the electron operator $c$ into partons and  \emph{define} symmetry transformations of parton operators. Furthermore, it is crucial to investigate and determine restrictions on the  values of $\theta$ if the gauge fields fluctuate strongly, which in parton constructions are assumed to be weak.
Weak gauge fluctuations \cite{maciejkoFTI,maciejko_model,maciejko2015,swingle2011} can be treated perturbatively, leading to reliable parton constructions. Once gauge fields fluctuate strongly, perturbation theory fails and the particle content is changed drastically. A  subtle treatment of this problem via 't Hooft's oblique confinement was given recently \cite{ye16a}. 

In this paper,    we derive    universal properties   of      QED$_4$ and   FTIs   \emph{on general grounds}.   
In Sec.~\ref{section_1},  we study the $S$-duality transformations of QED$_4$ in the presence of electron fractionalization. The modular parameter $\tau$ transforms under $SL(2,\Z)$ operation conjugated by a ``dilation'' operation, denoted by $\mathcal{D}^{-1} (SL)\mathcal{D}$. Here, $\mathcal{D}$ is the dilation operation on the complex plane and $SL$ represents elements of the $SL(2,\Z)$ group \cite{Witten1995,mo}. These combined transformations lead to the  ``fractional $S$-duality'': 
 \begin{align}
	 \mathcal{S}:\,\, \tau\rightarrow -\frac{1}{t^4\tau}\,\,;\,\,~~~~~~~~\mathcal{T}:\,\,\tau\rightarrow \tau+\frac{1}{t^2}\,,\label{eq_sec_intro_1}
 \end{align}
where  $t$ is a postiive integer and is related to the   fractionalized electric charge  carried by bulk excitations: $\pm {1}/{t}\,,\pm {2}/{t}\,,\cdots
$. Once $\mathsf{T}$  is imposed, the  fractional $S$-duality leads to a sequence of axion angles for FTIs:
\begin{align}
\theta=\frac{\pi}{t^2}\text{  mod } \frac{2\pi}{t^2}\,\,\label{quantized_axion_angle_main}
\end{align}  
which reduces to the familiar result $\theta=\pi \text{  mod }2\pi$ of TIs once $t=1$ \cite{Qi2008}.   In Sec.~\ref{section_2}, to better understand the microscopic mechanism of the fractional axion angles in Eq.~(\ref{quantized_axion_angle_main}), we propose a topological quantum field theory (TQFT) description of FTIs. Bulk fractional charge and fractional axion angle can arise from the deconfined phase \cite{fradkin_shenker,fradkin2013field} of a $\Z_N$ discrete gauge theory, where  $\text{U(1)}$   symmetry and $\mathsf{T}$  are imposed in a non-trivial way. 
This construction of FTIs to some extent generalizes the Chern-Simons field theory used in two space dimensions \cite{lopez_fradkin}. 
In Sec.~\ref{stacking_and_surface},  we consider the stacking operation \cite{string8} among  TI and FTIs.
By ``stacking'' we mean that two decoupled topological phases are put in the same spacetime region,  which leads to a new topological phase. In our case, each FTI is labeled by a minimal charge $\mathcal{Q}_m= {1}/{t}$ and an axion angle $\theta$.  We derive how $\mathcal{Q}_m$ and $\theta$ should add up properly under the stacking operation. 
From the analysis of the stacking, we find that FTIs can be divided into \emph{type-I FTIs} and \emph{type-II FTIs}. Type-I FTIs ($t\in\mathbb{Z}_{\rm odd}$)  can be realized by simply  stacking a TI \cite{TI4,TI6} and a $\mathbb{Z}_N$ topological phase with $\mathcal{Q}_m=1/t$ and a vanishing axion angle. In contrast, type-II FTIs ($t\in\mathbb{Z}_{\rm even}$)   cannot be realized in this way, and therefore represent a new class of FTIs that are  distinct from type-I. In Sec.~\ref{stacking_and_surface}, we also study the Surface Topological Order (STO) of both fermionic and bosonic FTIs. Our conclusions are discussed in Sec.~\ref{section_conclusion}

\section{{Fractional $S$-duality and dyon spectrum}}\label{section_1}

\subsection{Fractional $S$-duality}
 Consider a 3D fermionic gapped topological phase with a $\text{U(1)}$ global symmetry.  We add a background electromagnetic field $A_{\mu}$ that minimally couples to the fermionic current. Since the fermions are charged under the $\text{U(1)}$ global symmetry, we demand that the gauge field $A_\mu$ is a spin$_c$ connection that satisfies:  
\begin{align}
\frac{1}{2\pi}\int_\mathbb{U} F=\Z+\frac{1}{2}\int_\mathbb{U} \mathfrak{w}_2\,,
\end{align}  where $\Z$ denotes an arbitrary integer and $\mathbb{U}$ is a 2D closed sub-manifold.  $\mathfrak{w}_2$ is the second Stiefel-Whitney class \cite{nakahara_book} of the 4D spacetime manifold.  
By integrating out electrons, we obtain an effective action that takes exactly the same form as Eq.~(\ref{intro_action_main}).  However, due to the presence of fractional charges $\mathcal{Q}_m=1/t<1$ in the bulk, a stronger quantization of the integral $\frac{1}{2\pi}\int_\mathbb{U} F$ needs to be enforced.  
According to the famous Schwinger-Zwanziger quantization condition, two dyons (particle with electric charge $q$ and magnetic charge $p$) labeled by $(q, p)$ and $(q',p')$, respectively, must obey $qp'-q'p\in\Z$. Let us consider a system whose electrically charged excitations have a minimal charge $1/t$, i.e., there exists a particle with $q=1/t, p=0$. Then, for any dyon $(q',p')$, its magnetic charge $p'$ should be quantized: $p'=0, \pm t,\pm 2t,\cdots$, implying the quantization of the following integral: 
\begin{align}
 \int_{\mathbb{S}^2}F= 2\pi t \times \Z
 \end{align}
  with  $\mathbb{S}^2$ being a two-sphere ($\mathfrak{w}_2=0$). $\mathbb{S}^2$ spatially encloses  monopoles where $2\pi t \Z$ fluxes end. This stronger   condition was also observed in  superstring models \cite{wen_witten}. 
 
For generic $t$, we ask: what are the allowed duality transformations?   For this purpose, we  perform the  rescaling: $\tilde{g}={g}/{t}\,, \tilde{A}={A}/{t}\,, \tilde\theta=\theta t^2\,$, and obtain the  rescaled action 
\begin{align}
S=i\frac{\tilde \theta}{8\pi^2}\int \tilde F \wedge  \tilde F+\frac{1}{2\tilde g^2}\int \tilde F\wedge \star \tilde F,
\end{align}
 and a rescaled Dirac quantization condition $\frac{1}{2\pi}\int_{\mathbb{S}^2} \tilde F=\Z$.  Now, the theory is formally as same as QED$_4$ with integer charge. At this intermediate step, we may define a new modular parameter $\tilde{\tau}$:
$ \tilde\tau=\frac{\tilde\theta}{2\pi}+i\frac{2\pi}{\tilde{g}^2}=t^2\tau
 $ which must satisfy the following two standard  $SL(2,\Z)$ transformations \cite{Witten1995}:
$ \tilde \tau\longrightarrow -\frac{1}{\tilde \tau}\,$ and $\tilde\tau\longrightarrow \tilde \tau+1\,,$  
 where the shift by 1 is derived by equipping the spin$_c$ structure of the spacetime manifold. 
 These two transformations generate the $SL(2,\Z)$ group:
$\tilde \tau\rightarrow \frac{a\tilde \tau+b}{c\tilde\tau+d}\,, 
$ where the  integer numbers $a,b,c,d$ form   a unimodular matrix $
    \left(\begin{smallmatrix}a   &b\\ c&d\end{smallmatrix}\right)$. 
	By noting that $\tilde \tau=t^2\tau$, we obtain the $\mathcal{S}$ and $\mathcal{T}$-transformations in Eq.~(\ref{eq_sec_intro_1}). A direct consequence of the $S$-duality is the \emph{fractionalized} version of $\mathbb{Z}_2$ electromagnetic (strong-weak) duality for $\theta=0$: 
 \begin{align}
 g \, g^{dual} = {2\pi t^2}\,, 
 \end{align} 
where $g^{dual}$ is the coupling after the duality. When $t=1$, this $\mathbb{Z}_2$ strong-weak duality reduces to the standard one: $gg^{dual}=2\pi$ \cite{Witten1995,VAFA19943,mo}. Symbolically, the fractional $S$-duality can be written as: 
$ \mathcal{D}^{-1} ( SL) \mathcal{D}$,
 where $SL$ denotes  standard $SL(2,\Z)$ transformation and $\mathcal{D}$ denotes a  dilation transformation (multiplication by $t^2$).

 The fractional $\mathcal{T}$-transformation in Eq.~(\ref{eq_sec_intro_1})  is:
 \begin{align}
 \theta\longrightarrow \theta+\frac{2\pi}{t^2}
  \label{equation_theta_period}
 \end{align}
 which gives a $t$-dependent period $\Delta_\theta={2\pi}/{t^2}$.

 \subsection{Fractional modular fixed points} 
 
 With $\mathsf{T}$, $\theta$ is quantized at  ${\Delta_\theta}/{2}$ for FTIs, leading to Eq.~(\ref{quantized_axion_angle_main}) for FTI$_t$ (an FTI with minimal charge $1/t$). If $\theta=0$, the corresponding fractionalized state is denoted by TO$_t$.  In the presence of a boundary, since the bulk $F\wedge F$ topological response   reduces to a surface Chern-Simons term, we see that the surface has a ``fractionalized'' parity anomaly.  Moreover, since $ \mathcal{S}^2=(\mathcal{ST})^3=1$ even for $t\neq 1$,  all  fixed points \cite{fradkin_kivelson_96,modular_book} of $\tau$ are equivalent (under the fractional $S$-duality) to: ${i}/{t^2}$ and ${e^{i\frac{2\pi}{3}}}/{t^2}$. The former is   invariant under  the operation of $\mathcal{S}$ while the latter is  invariant under  the operation of $\mathcal{ST}$. 
While these modular fixed points with $t=1$ are physically relevant to  the phase diagram of the  TI surface after bosonization \cite{atma,fradkin_kivelson_96}, those with $t\neq 1$  may also lead to   surface critical points of FTIs. The details along this line is left to future work. 

  \subsection{``Integer'' and ``fractional'' gravitational response}
 Eq.~(\ref{intro_action_main})  is defined  on a flat spacetime  that admits spin$_c$ structure. However, in a curved spacetime, the   action should incorporate a contribution from the Riemann curvature $R$: 
 \begin{align}S=i\frac{\theta}{8\pi^2}\left(\int F \wedge  F-\frac{\sigma}{8}\right)+\frac{1}{2g^2}\int F\wedge \star F
 \end{align}
  as required by the {Atiyah-Singer} index  theorem. Here, $\sigma$ is the signature of the  manifold:
  \begin{align}\sigma=\frac{1}{48}\int\frac{\text{Tr}R\wedge R}{(2\pi)^2}\,.
  \end{align} 
 {In the fractional topological insulator states we studied, the gravitational effect is neglected for the sake of simplicity. It will be interesting to explore the possible formula in the presence of fractional axion angle}.

 \subsection{``Fractionalized'' parity anomaly on the surface of FTIs}

In addition to the \emph{fractional} $S$-duality and Witten effect, we may also introduce the notion of the \emph{fractional} parity anomaly on the 2D surface of FTIs. It is known that the usual TI surface hosts a single massless Dirac fermion that has a parity anomaly \cite{anom1,witten_RMP,anom4}. Classically, the action of a single Dirac cone is $\mathsf{T}$ invariant. However, when one proceeds to compute the effective action of $A$ on the surface, there do not exist regularization scheme such that the $\mathsf{T}$  and gauge invariance are preserved simultaneously. In the Pauli-Villars regularization, the single Dirac cone gives rise to an induced Chern-Simons term for $A$ at  half-level, which spontaneously breaks $\mathsf{T}$  on the surface. This Chern-Simons term can also be understood as a boundary term of the bulk axion term in Eq.~(\ref{intro_action_main}) with $\theta=\pi$:
\begin{align}\int i\frac{\theta}{8\pi^2}F\wedge F \rightarrow \int i\frac{1}{4\pi}\frac{1}{2}A\wedge d A \,.
\end{align} Likewise, on the surface of FTIs, we expect there is a fractionalized counterpart of the usual parity anomaly phenomenon \cite{anom1,witten_RMP,anom4}. Despite the lack of surface microscopic model, we may also indirectly describe the parity anomaly by dimensionally reducing the bulk axion term with $\theta$ given by Eq.~(\ref{quantized_axion_angle_main}):
\begin{align}\int i\frac{\theta}{8\pi^2}F\wedge F \rightarrow \int i\frac{1}{4\pi}\frac{1}{2t^2}A\wedge d A    \,.
\end{align}
 Therefore, the signal of the fractional parity anomaly on the surface of FTIs is given by the induced Chern-Simons term with the level quantized at $\frac{1}{2t^2}=\frac{1}{8},\frac{1}{18},\frac{1}{32},\frac{1}{125}\cdots$.

 \subsection{Understanding on $\Delta_\theta$ via dyon spectrum}
 \label{dyon_section}
 
 The $t$-dependent  $\Delta_\theta$ can also be understood from the  dyon spectrum. Before we analyze them, it is helpful to first classify the point-like excitations in the bulk once the $\text{U(1)}$ global symmetry is fully gauged. There are charged excitations, including electrons (charge $1$) and fractionally charged particles (carrying multiples of $1/t$ charge). It is also possible to have neutral nontrivial excitations in the bulk. Generally speaking, a dyon is a point-like object composed by externally imposed monopoles and bulk excitations. Denote the magnetic charge of a dyon by $m\in\Z$ (measured in unit of $2\pi t$), and the total number of electric charges (measured in unit of $1/t$) attached to the monopole by $n\in\Z$. There can also be neutral nontrivial excitations (that can be either bosonic or fermionic) attached to the monopole. We denote the total self statistical angle of all attached particles by $\Theta$.

 Then, we   characterize dyons with two quantum numbers: electric charge and self statistics. In analogy to the Witten effect in the gauged TI \cite{Qi2008,witten1} where $t=1$, the electric charge of a dyon is given by $\frac{n}{t}+\frac{\theta}{2\pi}tm$.  The self statistics of the dyon is then given by $e^{i(\pi mn+\Theta)}$ where $e^{i\pi mn}$ arises from the relative orbital motion between monopole and attached charged particles \cite{f1}.  Under the shift of Eq.\eqref{equation_theta_period}, the charge of a dyon is shifted by $\frac{m}{t}$. This change can be compensated by subtracting $m/t$ fractional charges via the shift $n\rightarrow n-m$. To maintain the statistics, these fractional charges should have self statistics $e^{i\Theta}=(-1)^m$.  In particular, if we set $m=1$ it implies that {the system \emph{must} allow a charge-$1/t$ fermionic excitation}. Although not entirely obvious, one can actually show that this is always the case. In fact, if we start with a charge-$1/t$ boson, then by fusing $t$ of them we obtain a charge-$1$ boson, whose bound state with an electron (more precisely, a hole) is a neutral fermion. Then, this neutral fermion and a charge-$1/t$ boson form a bound state which is a charge-$1/t$ fermion.  
  Therefore, in the fermionic topological phases that  we discussed so far, there must be  charge-$1/t$ fermionic excitations and the period $\Delta_\theta$ is always given by ${2\pi}/{t^2}$ (see Table~\ref{tab:period_appendix_fermion}).  
  
    \begin{table}[h]
\caption{$\text{U(1)}$-symmetric fermionic topological phases. For $\Delta_\theta$, there is only one choice, i.e., $\frac{2\pi}{t^2}$.  ``$1$'' means existence while ``$0$'' means that the corresponding quasiparticles do not exist.}
\begin{tabular}{c c c c c}
\hline
\hline
\begin{minipage}   {1.4in}  ~\\charge-$1/t$ boson\\~\end{minipage}  & \begin{minipage}   {0.9in}  0\end{minipage} & \begin{minipage}   {0.9in}1\end{minipage} \\\hline
\begin{minipage}   {1.4in} ~\\charge-$1/t$ fermion\\~\end{minipage}   & 1 &1 \\\hline
\begin{minipage}   {1.4in} ~\\neutral fermion\\~\end{minipage}   & 0 &1 \\\hline
\begin{minipage}   {1.4in} ~\\${\Delta_\theta}$\\~\end{minipage}  & ${2\pi}/{t^2}$ &$2\pi/{t^2}$  \\\hline\hline
\end{tabular}
\label{tab:period_appendix_fermion}
\end{table}%

\subsection{Bosonic topological phases with  $\text{U(1)}$ symmetry}

It is worth noticing that the above argument based on dyon spectrum can be applied to bosonic systems without much modification.  We therefore divide $\text{U(1)}$-symmetric bosonic topological phases further into two classes type-$\beta$, type-$\alpha$ (see Table~\ref{tab:period}), depending on whether or not there is a charge-$1/t$ fermionic quasiparticle in the bulk. Once $\mathsf{T}$ is imposed, we can define analogously bosonic FTIs (bFTI). It is easy to see that $\Delta_\theta=4\pi/t^2$ for type-$\alpha$ bFTIs and $\Delta_\theta=2\pi/t^2$ for type-$\beta$.
 We present details below.

  \begin{table}[h]
  \centering
\caption{Different types of $\mathcal{T}$-transformations 
 in $\text{QED}_4$ obtained by gauging the $\text{U(1)}$ symmetry of a topological phase of matter. $\Delta_\theta$ denotes the period of the axion angle $\theta$. 
}
\begin{ruledtabular}
\begin{tabular}{cccc}
 \begin{minipage}[c]{0.6in}Topological Phases\end{minipage} & \begin{minipage}[c]{0.8in}Fermionic Topological Phases\end{minipage}     & \begin{minipage}[c]{0.9in} Type-$\alpha$ Bosonic Topological Phases\end{minipage}     & \begin{minipage} [c]  {0.9in} Type-$\beta$ Bosonic Topological Phases \end{minipage}       \\ \hline
\begin{minipage}  [c] {0.5in} $ {\Delta_\theta}$\end{minipage}  & \begin{minipage}  [c] {0.5in}$ {2\pi}/{t^2}$\end{minipage}& \begin{minipage}[c]{0.5in}${4\pi}/{t^2}$\end{minipage}&\begin{minipage}  [c] {0.5in} ${2\pi}/{t^2}$\end{minipage} \\  
\end{tabular}
\end{ruledtabular}
\label{tab:period}
\end{table}%

  \begin{table}[h]
\caption{Two types of $\text{U(1)}$-symmetric bosonic topological phases. In this table, all possible bosonic topological phases are exhausted.  Once a quasiparticle exist, we assume its anti-particle also exists in the bulk spectrum.}
\begin{tabular}{c c c c c}
\hline
\hline
\begin{minipage}   {1.3in}  ~\\charge-$1/t$ boson\\~\end{minipage}  & \begin{minipage}   {0.6in}  1\end{minipage} & \begin{minipage}   {0.6in}0\end{minipage} & \begin{minipage}   {0.6in}1\end{minipage}\\\hline
\begin{minipage}   {1.3in} ~\\charge-$1/t$ fermion\\~\end{minipage}   & 0 &1 &1\\\hline
\begin{minipage}   {1.3in} ~\\neutral fermion\\~\end{minipage}   & 0 &0 &1\\\hline
\begin{minipage}   {1.3in} ~\\$ {\Delta_\theta}$\\~\end{minipage}  & ${4\pi}/{t^2}$ &${2\pi}/{t^2}$ &${2\pi}/{t^2}$ \\\hline
\begin{minipage}   {1.3in}  ~\\Types\\~\end{minipage}   & Type-$\alpha$ &Type-$\beta$&Type-$\beta$  \\\hline\hline
\end{tabular}
\label{tab:period_appendix}
\end{table}%

The period of $\theta$ can be determined by examining the self-statistics and electric charge of a dyon in gauged bosonic topological phases. The electric charge is given by:
$\frac{n}{t}+\frac{\theta}{2\pi} tm\,,
$ where the magnetic charge of a dyon is given by $m\in\Z$ (measured in unit of $2\pi t$), and $n\in\Z$ is the total number of electric charges (measured in unit of $1/t$) attached to the monopole. The self-statistics of a dyon is given by:
$ e^{i\pi mn+i\Theta}\,,
 $ where $e^{i\Theta}$ is determined by the  self-statistics (total fermion sign) of attached fermions (including both charged fermions and neutral fermions).  For type-$\alpha$ cases, all quasiparticles are bosonic, so $e^{i\Theta}=1$. In order to keep electric charge and self-statistics invariant under the period shift $\theta\rightarrow \theta+\Delta_\theta$,  $\Delta_\theta$ should be $\frac{4\pi}{t^2}$ and a shift in $n$ should be performed: $n\rightarrow n-2m$. For type-$\beta$ cases, $e^{i\Theta}$ can be either $1$ or $-1$ due to attached fermions. Therefore,  the derivation is similar to that of fermionic topological phases in Sec.~\ref{dyon_section}. The result is $\Delta_\theta=\frac{2\pi}{t^2}$. As  a result, we end up with Table~\ref{tab:period} and Table~\ref{tab:period_appendix}.
 
 By further taking $\mathsf{T}$ into consideration, the $\mathsf{T}$ invariant points of $\theta$ is either $0$ or half of the period. The former corresponds to bTO$_t$ (in analogy to TO$_t$) with two types ($\alpha$bTO$_t$ and $\beta$bTO$_t$); the latter corresponds to bFTI$_t$ (in analogy to FTI$_t$) with two types ($\alpha$bFTI$_t$ and $\beta$bFTI$_t$).
  In Table~\ref{tab:period_boson}, we list all bosonic topological phases with $\text{U(1)}$ symmetry and $\mathsf{T}$. ``bTI'' denotes the bosonic topological insulator\cite{lapa17,bti2,bti1}, which is the bosonic analogue of the usual fermionic topological insulator (TI). ``bVac'' denotes a trivial bosonic Mott insulator with symmetries where the \emph{only} excitations are trivial particles (charge-1 bosons). In comparison, ``Vac''  that will be introduced in Sec.~\ref{stacking_and_surface}  denotes a trivial band insulator where  the \emph{only} excitations are trivial particles (charge-1 fermions). ``bTO'' is the bosonic analogue of ``TO'' that will be introduced in Sec.~\ref{stacking_and_surface}. ``$\alpha$bTO'' and ``$\beta$bTO'' are two types of bTO with different $\Delta_\theta$ values as shown in Table~\ref{tab:period_appendix}. ``bFTI'' is the bosonic analogue of ``FTI''. ``$\alpha$bFTI'' and ``$\beta$bFTI'' are two types of bFTI with different $\Delta_\theta$ values as shown in Table~\ref{tab:period_appendix}.
 
    \begin{table}[h]
\caption{All bosonic topological phases with $\text{U(1)}$ symmetry and $\mathsf{T}$. The surface topological order of $\alpha$bFTI$_t$ with $t=N$ is studied in Sec.~\ref{STO_BFTI}.}
\begin{tabular}{c c c c c c c}
\hline
\hline
\begin{minipage}   {0.5in}  ~\\~\\~\end{minipage}  & \begin{minipage}   {0.3in}  bTI\end{minipage} & \begin{minipage}   {0.3in}  bVac\end{minipage} & \begin{minipage}   {0.4in}  $\alpha$bTO$_t$\end{minipage} & \begin{minipage}   {0.4in}  $\beta$bTO$_t$\end{minipage} & \begin{minipage}   {0.5in}  $\alpha$bFTI$_t$\end{minipage} & \begin{minipage}   {0.5in}  $\beta$bFTI$_t$\end{minipage} \\\hline
\begin{minipage}   {0.5in}  ~\\${\Delta_\theta}$\\~\end{minipage}  & \begin{minipage}   {0.3in}  $4\pi$\end{minipage} & \begin{minipage}   {0.3in}  $4\pi$\end{minipage} & \begin{minipage}   {0.4in}  ${4\pi}/{t^2}$\end{minipage} & \begin{minipage}   {0.4in}  ${2\pi}/{t^2}$\end{minipage} & \begin{minipage}   {0.4in}   ${4\pi}/{t^2}$\end{minipage} & \begin{minipage}   {0.4in}   ${2\pi}/{t^2}$\end{minipage} \\\hline
\begin{minipage}   {0.7in}  ~\\$\theta$ in the first ``Brillouin zone''\\~\end{minipage}  & \begin{minipage}   {0.3in}  $2\pi$\end{minipage} & \begin{minipage}   {0.3in}  $0$\end{minipage} & \begin{minipage}   {0.4in}  $0$\end{minipage} & \begin{minipage}   {0.4in}  $0$\end{minipage} & \begin{minipage}   {0.4in}  ${2\pi}/{t^2}$\end{minipage} & \begin{minipage}   {0.4in}  ${\pi}/{t^2}$\end{minipage} \\\hline
\hline
\end{tabular}
\label{tab:period_boson}
\end{table}%

  \section{{Topological quantum field theory}}
  \label{section_2}
  
   Below we propose a family of FTIs with TQFTs, which provides a ``microscopic'' justification of the sequence of fractional axion angles and the fractional $S$-duality.  
We start with a  $\Z_N$ gauge theory in the deconfined phase \cite{fradkin_shenker} whose low-energy effective action is a $BF$ theory \cite{horowitz}:
\begin{align}
S=i\int \frac{N}{2\pi} b\wedge da+S_{ m}+i\int j\wedge \star a+i\int \sigma\wedge \star b\,,
\label{action_bf_theory}
\end{align}
where $b$ and $a$ are 2-form and 1-form  gauge fields respectively.  The Maxwell term  $S_m$ is given by: 
\begin{align}S_m=\frac{1}{2g^2_a}\int da\wedge \star da+\frac{1}{2g_0^2}\int F\wedge \star F\,,
\end{align}  where $g_a$ and $g_0$ are the bare gauge couplings of $a_\mu$ and $A_\mu$ respectively. The gauge coupling of $b_{\mu\nu}$ has been dropped since it  is irrelevant under the renormalization group flow. The bulk topological order \cite{xgwen} can be understood via this $\mathbb{Z}_N$ gauge theory. Quasiparticle excitations can be understood as gauge charges of the  $\mathbb{Z}_N$ gauge group. $j$ and $\sigma$ are 1-form world-line of quasiparticles (carrying unit gauge charge of $a$) and 2-form world-sheet of loops (carrying unit gauge charge of $b$) respectively. The gauge transformations are defined as:
$ b\rightarrow b+d V\,,\,a\rightarrow a+d\lambda  
$\,,
where $V$ and $\lambda$ are 1-form and 0-form gauge parameters respectively. Gauge invariance in a closed and compact Euclidean manifold requires that: $
N\in\Z\,,\,d\!\star \! \wedge j=0\,,\, d\!\star\! \wedge\sigma=0
$.   Eq.~(\ref{action_bf_theory}) is the minimal field-theoretic model as shown below.   Indeed, one may  consider  {twisted}   gauge theories \cite{bti2_prb,3dset_plus,3dset_twisted,kapustin,wang1,wang2,hehuan,chenxiao_2016,chan_ye_ryu} with new terms (e.g., $a^1\wedge a^2\wedge da^2$ with $a^{1},a^2,\cdots$ being  1-form gauge fields), leading   to more exotic FTIs. The TQFT as given describes a bosonic system since no spin$_c$ structure is required to make sense of the theory, however we assume that trivial complex fermionic excitations exist (i.e. forming a trivial band insulator) so the full theory is well-defined only on spin$_c$ manifolds.

Next, we impose $\text{U(1)}$ and $\mathsf{T}$ symmetries. 
The action of $\text{U(1)}$ symmetry can be defined by coupling to the electromagnetic field $A$ ($F=dA$):
\begin{align}
S_\text{coupling}=-i\int \frac{r}{2\pi}b\wedge F+i\int \frac{s}{2\pi} da\wedge F\,,\label{eq_label_sym}
\end{align}
where $r\in\Z$ required by gauge invariance.    {In Ref.~\onlinecite{comment_paper},   an action is proposed that is a special case of Eq.~(\ref{eq_label_sym}) with $r=1$ (therefore, $t=N=K$). Using their notation [see their Eq.~(3.4)], the fractional axion angle  is $\pi/N$, which is  different from our result $\pi/N^2$.  Their $\theta$ (equal to $4\pi s$)  takes   $\pi$, which is too restrictive to achieve the minimal value ${\pi}/{N^2}$ that we found here. 

   In principle, $s$ is allowed to take arbitrary real values if $\mathsf{T}$  is not required.  The  term ``$-\frac{r}{2\pi}b\wedge F$'' implies that an elementary gauge charge (represented by   $j$) carries $\frac{r}{N}$ $\text{U(1)}$ charge. To see this, we use equation of motion to obtain $Na=rA$. Then we have the coupling $j\wedge \star a=\frac{r}{N}j\wedge \star A$. A general particle excitation can be formed by $k$ units of gauge charge $k$ and $k'$ electrons, which then carries $\frac{kr+k'N}{N}$ $\text{U(1)}$ charges. Since $kr+k'N$ is always a multiple of $\mathsf{Gcd}(N,r)$,  the minimal charge is $1/t$, where
\begin{align}t=\frac{N}{\mathsf{Gcd}(N,r)}.
\end{align}  The on-shell condition $Na=rA$ also implies that the Dirac quantization for $F$ becomes
$	r\int_{\mathbb{S}^2}F= N\int_{\mathbb{S}^2}da=N\cdot 2\pi\mathbb{Z}.
$ Thus $\int_{\mathbb{S}^2}F=\frac{2\pi N}{r}\mathbb{Z}$. In addition, $F$ is constrained by the usual Dirac condition of the 1-form gauge field $A$: $\int_{\mathbb{S}^2} F=2\pi \mathbb{Z}$.
In order to satisfy both conditions, the solution takes the form of:  
\begin{align}	\int_{\mathbb{S}^2} F=2\pi t\mathbb{Z}.
\end{align}

Once we integrate out the bulk gauge fields (or, simply insert the saddle point solution from the equation of motion since the theory is quadratic), we obtain Eq.~(\ref{intro_action_main}), where the axion angle is parametrized by $r$, $s$, and $N$:
$
\theta = {4\pi rs}/{N}\,,
$
and, 
 $g$ is determined by
 \begin{align}g= \frac{1}{\sqrt{{r^2}g_a^{-2} N^{-2}+g_0^{-2}}}.
 \end{align} 
 In analogy to Ref.~\onlinecite{Witten1995} but with   different inserted fluxes, we may evaluate the $\theta$ term (denoted by $S_{\theta}$) on  $\mathbb{M}_1\times \mathbb{M}_2$ (both $\mathbb{M}_1$ and $\mathbb{M}_2$ are two-spheres $\mathbb{S}^2$) and the field configuration: $\int_{\mathbb{M}_1} F_{12}=\int_{\mathbb{M}_2} F_{34}=2\pi t$.  As a result, $S_{\theta}=i\theta t^2$. Therefore, if we perform the period shift $\theta\rightarrow \theta+{2\pi}/{ t^2}$, the total action is changed by $2\pi$, which leaves the partition function unaltered. This period shift also corresponds to the shift of the parameter $s$:  
\begin{align}s\rightarrow s+\frac{\mathsf{Gcd}(N,r)}{2\mathsf{Lcm}(N,r)}\,,
\end{align}
where $\mathsf{Lcm}(N,r)$ denotes the least common multiple of $N$ and $r$. If $\mathsf{T}$-symmetry  is further imposed, $a$ and $A$ transform as polar-vectors  (i.e., $a_{0} \rightarrow a_{0} $, $a_{i} \rightarrow -a_{i} $ with $i=1,2,3$, so does $A$) while $b$ transforms as an axial-vector (i.e., $b_{0i}\rightarrow -b_{0i}$, $b_{ij}\rightarrow b_{ij}$ with $i,j=1,2,3$). The $s$-dependent term in Eq.~(\ref{eq_label_sym}) changes sign but $s$ can be quantized at:
\begin{align}s=\frac{\mathsf{Gcd}(N,r)}{4\mathsf{Lcm}(N,r)} \text{ mod }\frac{\mathsf{Gcd}(N,r)}{2\mathsf{Lcm}(N,r)}   
\end{align} such that the period shift of $s$ recovers $\mathsf{T}$. Meanwhile, this quantization of $s$ leads to the quantization of $\theta$ in Eq.~(\ref{quantized_axion_angle_main}).

\section{Stacking operation and surface topological order}
\label{stacking_and_surface}

\subsection{Classification of FTIs via stacking operation}

 We consider a simple operation that stacks FTIs together to get a new FTI.  One reason to consider stacking, as we will see, is to distinguish two types of FTIs: {type-I FTI} and {type-II FTI}.  ``Stacking'' means that the resulting state is described by a new Hamiltonian operator that is formed by simply taking tensor product of   two original Hamiltonians. Mathematically, the stacking operation is an associative binary operation and the trivial band insulator (to be denoted by ``Vac'') is the identity element. All topological phases connected via stacking  form a monoid \cite{string8} in which \emph{only} TI has an inverse element (i.e., itself). 

We  label an FTI by ``FTI$_{t}$'' where $t$ specifies the minimal charge $1/t$ as well as $\theta$ in Eq.~(\ref{quantized_axion_angle_main}). Considering stacking two FTIs together (symbolically denoted by ``$\text{FTI}_{t_1} \boxtimes\text{FTI}_{t_2}$'').  In the stacked phase, it is easy to see that the minimal charge is ${1}/{t_*}$ with
\begin{align}
t_*=\mathsf{Lcm}(t_1,t_2)\,. 
\end{align}
 The axion angle is additive: 
\begin{align}
\theta_*= (\frac{\pi}{t_1^2}+\frac{\pi}{t_2^2} ) \text{  mod  }\frac{2\pi}{t^2_*}\,
\end{align}
 since both phases before stacking respect $\text{U(1)}$ global symmetry and are coupled to electromagnetic field $A$. 
 Clearly, $-\theta_*$ can always be connected to $\theta_*$ through a proper period shift, which is consistent with the $\mathsf{T}$.  If  $\theta_*$ cannot be connected to zero, the stacked phase is a new FTI with $\theta_*=\frac{\pi}{t^2_*}\text{   mod  }\frac{2\pi}{t^2_*}$.    Otherwise, the stacked phase does not exhibit nontrivial Witten effect (i.e., $\theta_*=0\text{ mod }\frac{2\pi}{t^2_*}$). We denote such a state by ``TO'' (topological order). In other words, the state TO$_t$ is topologically ordered with fractionalized charge excitations (minimal charge $\mathcal{Q}_m=1/t$) but its electromagnetic response is trivial in a sense that  the axion angle is vanishing (i.e., $\theta=0\text{ mod }\frac{2\pi}{t^2}$).   
Practically we may examine the parity of  the ratio: ${2 (\frac{\pi}{t_1^2}+\frac{\pi}{t_2^2} )}/ ({\frac{2\pi}{t^2_*}} )$ which has the same parity of $\frac{t_*}{t_1}+\frac{t_*}{t_2}$. Therefore, if $\frac{t_*}{t_1}+\frac{t_*}{t_2}$ is even, the new phase is a TO$_{t_*}$; otherwise, it is an FTI$_{t_*}$.

Let us work out some examples. First, we note that if $t_1=t_2=1$, the stacking between two TIs is obtained: 
\begin{align}\mathrm{TI}\boxtimes\mathrm{TI}=\mathrm{Vac}\,,\label{ti_ti_vac}
\end{align}
where $\mathrm{Vac}$ denotes a trivial band insulator (i.e., a vacuum with trivial fermions) whose axion angle $\theta=0\text{ mod }2\pi$. This stacking equation reflects the well-known $\Z_2$ classification of TIs \cite{TI4,TI6}. It also reflects the fact that TI is \emph{invertible}: its inverse is itself. It is   straightforward to verify that:
\begin{align}\mathrm{TO}_{t_1}\boxtimes\mathrm{TO}_{t_2}=\mathrm{TO}_{t_*} .
\end{align} Then, let us consider $t_1=t>1, t_2=1$. The stacking equation is given by:
\begin{align}
\mathrm{FTI}_{t}\boxtimes\mathrm{TI}=
\begin{cases}
\mathrm{TO}_{t} & t\in\Z_{\rm odd}\,; \\
\mathrm{FTI}_{t} & t\in\Z_{\rm even}\,.
\end{cases}\label{eq:st}
\end{align} 
The stacking relation reveals an interesting feature of FTIs with odd $t$: such a state can be thought as a simple stacking of a TO$_t$ (with odd $t$) and a TI. This is, however, not true when $t$ is even.  By adding ``$\boxtimes \text{TI}$'' on both sides of Eq.~(\ref{eq:st}) and using  Eq.~(\ref{ti_ti_vac}), we may further confirm this observation: 
\begin{align}
\mathrm{TO}_{t}\boxtimes \mathrm{TI}=\mathrm{FTI}_t\,\,,\, (t\in\Z_{\rm odd})\,.
\end{align}  
Actually, we also have: 
\begin{align}
\mathrm{TO}_{t}\boxtimes\mathrm{TI}= \mathrm{TO}_{t}\,( t\in\Z_{\rm even})\,.\label{striking}
\end{align}

From the above analysis, it is meaningful to   classify all FTIs into two classes: odd $t$ ({type-I FTI}) and even $t$ ({type-II FTI}). 
  In order to gain a deeper insight on the underlying physics, it is beneficial to derive the rules for all other possible stacking.  For example, we first consider the following stacking: 
  \begin{equation}
	\mathrm{FTI}_{t_1}\boxtimes\mathrm{FTI}_{t_2} =
	\begin{cases}
		\mathrm{TO}_{t_*} & t_1, t_2\in \Z_{\rm odd}\,;\\
		\mathrm{FTI}_{t_*} & t_1\in \Z_{\rm odd}, t_2\in \Z_{\rm even}\,.
	\end{cases}
	  \label{}
  \end{equation}
 The first equation  simply indicates that any two type-I FTIs, after stacking, lead to a vanishing axion angle; The second equation indicates that a stacking of a type-I FTI and a type-II FTI always leads to a type-II FTI ($t_*\in\Z_{\rm even}$). It means that the effect of the type-I FTI$_{t_1}$ in this stacking is  equivalent to that of a TO$_{t_1}$.  

However,  if both $t_1$ and $t_2$ are even, the stacking relation is slightly more involved. Write $t_i=2^{k_i}t_i'$ for $i=1,2$ where $t_i' \in\Z_{\rm{odd}}$ and $k_i\in\Z$. Then,
  \begin{equation}
	 \mathrm{FTI}_{t_1}\boxtimes\mathrm{FTI}_{t_2} =
	\begin{cases}
		\mathrm{TO}_{t_*} & k_1= k_2\,;\\
		\mathrm{FTI}_{t_*} & k_1\neq k_2\,,
	\end{cases}
	  \label{}
  \end{equation}
  where 
  \begin{align}
  t_*=\mathsf{Lcm}(t_1',t_2')2^{\text{max}(k_1,k_2)}\,.
  \end{align}  In summary,  all stacking operations  among $\mathrm{TI}$,  $\mathrm{TO}_t$, type-I $\mathrm{FTI}_t$ and type-II $\mathrm{FTI}_t$ can be obtained. Practically, the  stacking among three classes of ``root phases'', namely, TI, $\mathrm{TO}_t$ ($t\in\Z_{\rm odd}$) and type-II $\mathrm{FTI}_t$, suffices to generate the whole monoid.   
  
Before ending the discussion of the stacking operation, it is also beneficial to list the stacking results for $\alpha$bFTIs  (see Tables \ref{tab:period_appendix} and \ref{tab:period_boson}).  The results are similar to FTIs discussed above. Each $\alpha$bFTI is labeled by $t$, which gives the value of $\theta$: $\theta=\frac{2\pi}{t^2}\text{ mod }\frac{4\pi}{t^2}$. The usual bosonic topological insulator (bTI)  corresponds to $\theta=2\pi \text{  mod  }4\pi$.  According to the stacking operation, we can divide all bFTIs into type-I (with odd $t$) and type-II (with even $t$).

\subsection{Trivializing the Moore-Read $\times$ anti-semion surface topological order}

The relation (\ref{striking}) has an interesting consequence: consider the boundary between a TI and a TO$_t$ with even $t$. The stacking relation implies that the boundary can be completely trivial. In fact, we can explicitly check that this is indeed the case for a particular surface termination.  Before we proceed, let us clarify that since the system is made out of electrons, charge-statistics relation implies that all local bosons have to carry even charges. Therefore, TO$_t$ with even $t$ contains a charge-$1$ bosonic particle, while no such particles exist for odd $t$.

Imagine driving the TI surface to the Moore-Read $\times$ anti-semion topological order~\cite{STO4, STO3}. As shown in Ref. \onlinecite{STO4}, condensing a charge-$1$ boson $e^{4i\phi}$ results in the so-called $\text{U(1)}$-breaking SC$^*$ phase, which is not anomalous.  Proximity to TO$_t$ implies that we can bring in a boson with $-1$ charge from the TO$_t$ bulk, and condense the bound state of $e^{4i\phi}$ (following the notation of Ref. \onlinecite{STO4}) with this boson, which is now charge-neutral. Therefore, the condensation preserves all symmetries now, and we can completely trivialize the surface. A similar phenomenon was observed in 2D FTIs as well~\cite{wanglevin}.

\subsection{Surface topological order of $\alpha$bFTIs}\label{STO_BFTI}

\subsubsection{Bulk construction via projective construction and dyon condensation}

In this subsection we discuss an example of surface topological order for bFTIs of type-$\alpha$, i.e., $\alpha$bFTIs (see Tables \ref{tab:period_appendix} and \ref{tab:period_boson}). In other words, all excitations are bosonic (after turning off the external electromagnetic field $A_\mu$). First we make several general remarks about gapped surface states in the presence of bulk topological order. Different from the case of SPT bulk, now we must distinguish two kinds of localized excitations that can appear on a gapped surface: first, there are truly surface anyons, which only move on the surface. Second, the bulk particles can also appear on the surface, and one can count them as part of the surface anyon spectrum as well. 
However, we should note that the bulk particles, when viewed in the surface theory, are ``transparent'', i.e. the braiding statistics with all other surface anyons are trivial.  Therefore they appear to be ``local'' when only surface excitations are concerned. We will therefore identify the surface topological order counting these bulk particles as ``surface  local excitations''.

 First we present a construction of the bulk $\alpha$bFTI phase, generalizing a similar construction for electron TI in Ref. \onlinecite{metlitski_duality}. The idea of the dyon condensation that will be used below can be  found in Ref. \onlinecite{ye16a,YW13a}. The starting point is a $\mathrm{U}(1)$ spin liquid in the bulk with a
gapless photon $a_\mu$ and fermionic partons $\psi_\alpha$ (charged under the $\mathrm{U}(1)$ gauge symmetry). Under
time-reversal symmetry, $\psi_\alpha$ transforms as:
$	\mathsf{T}:
\psi \rightarrow i\sigma^y \psi^\dagger. 
$ Namely the symmetry group of $\psi$ is $U(1)_\text{g} \times \mathbb{Z}_2^{T}$, where $U(1)_\text{g}$ is
the gauge group. The electric and magnetic field of $a_\mu$ are
odd and even under time-reversal respectively, which is opposite
to the transformation of the external U(1) gauge field $A_\mu$.

At the mean-field level, when gauge fluctuations are ignored, we assume that the fermions form a class AIII superconductor with $\nu=2$~\cite{schnyder2008, kitaev2009}. We can then include the gauge fluctuations and obtain the $\mathrm{U}(1)$ spin liquid. Since fermions are gapped, at very low energies there are only gapless photons. Due to the topological band structure, the effective action of $a_\mu$ has a topological term with $\theta=2\pi$.
 Due to the Witten effect of
the topological insulator, a $2\pi$-monopole of
$a_\mu$ will carry a polarization gauge charge
$\frac{\theta}{2\pi}= 1$. In general, a dyonic excitation in this spin liquid
can be labelled as $(q, m)$ where $q$ is the total gauge charge
and $m$ the monopole number. The Witten effect then implies $q =
n +  m$, where $n\in\mathbb{Z}$ is
the number of partons $\psi$ attached to the dyon. Thus a
$2\pi-$monopole can be neutralized by binding with a hole 
 of $\psi$. We label this neutralized monopole as the
$(0,1)$ monopole.

In the bulk we condense the bound state of a $(0,N)$ monopole
and a physical boson that carries no gauge charge but one global
U(1) charge. It should be noticed that $(0,N)$ monopoles have ${\mathsf{T}}^2=(-1)^N$~\cite{metlitski_TSC}.
So the physical boson must be a
Kramers doublet (singlet) under $\mathsf{T}$ if $N$ is odd (even). We also assume that it carries charge
$+1$ under the external gauge field $A_\mu$. This  bound
state is a gauge neutral, time-reversal invariant, and charge-1
boson.  We
can label all bulk point-like excitations in terms of their quantum numbers $(q,
m, Q, M)$, where $q$ is the gauge charge under $a_\mu$, $m$ is the
monopole number of $a_\mu$, $Q$ is the global U(1) symmetry
charge, and $M$ is the monopole number of the external U(1) gauge
field $A_\mu$. Under this notation, the condensed bound state has
quantum number $(0, N, 1, 0)$. 
The condensate of the aforementioned bound state $(0, N, 1,
0)$ will confine all the excitations that have nontrivial
statistics with it, including the $\psi$ fermions. What remains deconfined are monopoles with $2\pi k, k=0, 1,\dots, N-1$ fluxes, so we obtain a $\mathbb{Z}_N$ topological order.  The global U(1)
symmetry is still preserved because the condensed bound state is
also coupled to the dynamical gauge field $a_\mu$.  

If we move
a $2\pi M$ Dirac monopole of $A_\mu$ into the bulk, due to the Meissner effect of the dyon condensate, we must have $M=Nq$,
to avoid
confinement caused by the condensate of the $(0, N, 1, 0)$ bound
state. There are two physical implications: first, external monopoles are only allowed in multiples of $N$ to avoid confinement, which is consistent with charge fractionalization in the bulk. Second, a $A_\mu$ monopole of strength $N$ will automatically pair with a gauge
fermion $\psi$ (with $q=1$) to form a bound state with quantum number $(1, 0,
0, N)$, so it has trivial mutual statistics with the condensed
$(0, N, 1, 0)$ bound state. This Dirac monopole $(1,
0, 0, N)$ is neutral under the global U(1) symmetry, but it is a
fermion. This neutral fermionic Dirac monopole of the external
gauge field $A_\mu$ is the characteristic statistical Witten 
effect of the bosonic FTI state with U(1) and time-reversal
symmetry. In other words, we find that the axion angle is
\begin{equation}
	\theta=\frac{2\pi}{N^2} \text{ mod }\frac{4\pi}{N^2}.
	\label{}
\end{equation}

\subsubsection{Effective theory of surface topological order}

  Now we derive the
 surface theory of the system. The main steps are rather straightforward generalizations of those in Ref. \onlinecite{metlitski_duality}, so we will not go into details. After the confinement transition, the gauge field $a$ only exists on the surface, coupled to two surface Dirac cones. Since we
condense the bound state $(0,N,1,0)$ in the bulk, the surface must
have a $\frac{i}{2\pi N}\epsilon_{\mu\nu\lambda}a_\mu \partial_\nu
A_\lambda$ Chern-Simons term. In the end, we arrive at the following surface theory:
\begin{equation}
	\mathcal{L}=\sum_{j=1,2}\bar{\psi}_j\gamma^\mu(\partial_\mu-ia_\mu)\psi_j + \frac{i}{2\pi N}\epsilon_{\mu\nu\lambda}a_\mu \partial_\nu A_\lambda.
	\label{}
\end{equation}
Here $\psi_j$ is a Dirac fermion.
Notice that if the external gauge field $A$ is turned off, the low-energy theory is simply QED$_3$ with two flavors of Dirac fermions $\psi_j$. We shall not address the strong-coupling dynamics of the theory, since we will be interested in the massive deformation.

The surface can be driven into a gapped topological phase by turning on a $s$-wave pairing for Dirac fermions. To understand the gapped phase, we first need to analyze the properties of vortices in the (mean-field) superconducting state, which have been worked out in Ref. \onlinecite{metlitski_TSC} and we will briefly review below. First we suppress the dynamics of gauge fluctuations, and focus on the properties of vortices in the mean-field topological superconductor. We use the slab trick~\cite{metlitski_TSC}, and turn on a $s$-wave superconducting order parameter on the top surface, and $\mathsf{T}$-breaking mass on the bottom surface:
$	\delta H_\text{bottom}=m\sum_j \bar{\psi}_j\psi_j.$ 
The whole slab can be viewed as a topological superconductor with Chern number $2$, whose vortex statistics is described by $\mathrm{U}(1)_4$. Anyons in $\mathrm{U}(1)_4$ are labeled by a mod 4 integer $n$, whose self statistics is $e^{i\frac{\pi n^2}{4}}$. Notice that $n=1,3$ correspond to vortices, but $n$ itself is not the vorticity. $n=2$ is the Bogoliubov quasiparticle in the superconducting surface, which should have $\mathsf{T}^2=-1$ (in the mean-field theory).
 We also need to substract the contribution from the bottom surface. Let $k\in\mathbb{Z}$ denote the vorticity, the excitations are labeled by the pair $([n]_4,k)$ subject to the condition that $n+k$ is even, whose self statistics is $e^{\frac{i\pi (n^2-k^2)}{4}}$.
 Under time-reversal symmetry, we have $n\rightarrow -n$ mod 4, and $k$ is invariant. So $([n]_4, k)\rightarrow ([-n]_4, k)$. One can also show that even $k$ vortices are (ordinary) Kramers doublet~\cite{metlitski_TSC}. This is consistent with $(0,1)$ monopole having $\mathsf{T}^2=-1$, since when it passes through the surface a $hc/e$ vortex ($k=2$) is left behind. Furthermore, this particular $hc/e$ vortex has no mutual braiding statistics with any other vortices and Bogoliubov quasiparticles, so we can identify it with $(2,2)$ in the surface theory. In other words, $(2,2)$ is a surface local excitation, descending from a nontrivial particle in the bulk.
 Finally, due to the Chern-Simons term, vortices with $k$ vorticity carry charge $\frac{k}{2N}$ since they trap $\pi k$ flux of $a$. The $(2,2)$ particle then carries $\frac{1}{N}$ charge, consistent with the identification as the fundamental particle in the bulk $\mathbb{Z}_N$ topological order.

Next we take into account gauge fluctuations of $a_\mu$. With dynamical $a_\mu$ gauge field, the vortices have short-ranged interactions and the surface becomes fully gapped. The quasiparticles of the resulting topological order are equivalence classes of the $(n,k)$ moding out local excitations (including the bulk particles, as discussed earlier).  
We have identified that $(2,2)$ and $(0,4)$ are both bulk particles, so we only need to consider $n,k\in\{0,1,2,3\}$. It is not difficult to find the complete list of distinct anyons:
\begin{equation}
	\begin{split}
		(0,0) &\sim 1,\\
		(1,1) &\sim e,\\ 
		(1,3) &\sim m,\\
		(2,0) &\sim em.
	\end{split}
	\label{}
\end{equation}
One can easily check that the topological order is identical to that of a $\mathbb{Z}_2$ toric code (hence the labels). $e$ has $\frac{1}{2N}$ charge and $m$ has $\frac{3}{2N}$ charge. Formally, we can describe the surface topological order by a $\mathrm{U}(1)\times\mathrm{U}(1)$ Chern-Simons theory with the following K matrix and charge vector:
\begin{equation}
	K=
	\begin{pmatrix}
		0 & 2\\
		2 & 0
	\end{pmatrix},
	\mathbf{t}=\frac{1}{N}
	\begin{pmatrix}
		3\\
		1
	\end{pmatrix}.
	\label{}
\end{equation}
Therefore, in a strictly two-dimensional realization there must be a Hall conductance $\sigma_{xy}=\mathbf{t}^\mathsf{T}K^{-1}\mathbf{t}=\frac{3}{N^2}$, as expected.

We can also see the distinction between type-I and type-II $\alpha$bFTIs. When $N$ is odd, we have $\frac{1}{2N}=\frac{1}{2}-\frac{(N-1)/2}{N}$, which means that we can attach bulk particles to the surface anyons so that $e$ and $m$ anyons both have $1/2$ charge. Therefore, the surface is identical to the one of a bosonic TI (i.e., bTI), plus fractionalized charged particles (i.e., $\alpha$bTO) from the bulk. This is in complete agreement with the simple picture of type-I $\alpha$bFTI. A recent related work on surface topological order  can be found in Ref.~\onlinecite{2017arXiv170600429C}.

  \section{{Conclusions and Outlook}}
   \label{section_conclusion}
  
 In this paper, we proposed the concept of the fractional $S$-duality which generalizes the conventional $S$-duality in high energy physics and is useful in condensed matter physics. We also studied fractional topological insulators on a general ground and try to give a simple classification and characterization of fractional topological insulators by using topological quantum field theory and the stacking operation. Furthermore, we studied the surface topological order of fractional topological insulators. There are several potential directions for further investigation. One is to consider the fractional $S$-duality on a 4D manifold with boundary and to derive possible new types of dualities on (2+1)D \cite{vortexduality1}.   It will be beneficial to systematically study type-II FTIs (e.g., a   parton construction \cite{ye16a,YW13a,ye12,Ye14b}).
 Another direction is to define  FTI series in all even spacetime dimensions, similar to Ref.~\onlinecite{lapa17}. 
 
\begin{acknowledgments}
We would like to thank  M. F. Lapa and K. Shiozaki for useful discussions.  P.Y. acknowledges Z.-X. Liu's warm hospitality and helpful discussions at the Renmin University of China, with the support of the University's Research Fund (No.~15XNFL19), the NSF of China (No.~11574392), and the Major State Research Development Program of China (2016YFA0300500).  P.Y. also thanks Prof.~Tao Xiang for warm hospitality during the visit to the Institute of Physics, Chinese Academy of Sciences. P.Y. and E.F. were supported in part by the National Science Foundation through grant DMR~1408713 at the University of Illinois. 
\end{acknowledgments}


%

\end{document}